\def\rfr#1{eq. (\ref{#1})}

\def\asec{$''$ cy$^{-1}$}

\def\asec{$''$ cy$^{-1}$}

\def\bar{\begin{eqnarray}}
\def\ear{\end{eqnarray}}
\def\bb{\bibitem}
\def\eqi{\begin{equation}}
\def\eqf{\end{equation}}
\def\eqia{\begin{eqnarray}}
\def\eqfa{\end{eqnarray}}
\def\rp#1#2{{#1\over#2}}

\def\lb#1{\label{#1}}

\def\oc2{$\mathcal{O}(c^{-2})$}

\documentclass{aastex}
\usepackage{url}

\RequirePackage{color}

\begin{document}

\title{On the recently determined anomalous perihelion precession of Saturn}

\shorttitle{On the recently determined anomalous perihelion precession of Saturn}
\shortauthors{L. Iorio }

\author{Lorenzo Iorio }
\affil{INFN-Sezione di Pisa. Permanent address for correspondence: Viale Unit\`{a} di Italia 68, 70125, Bari (BA), Italy. E-mail: lorenzo.iorio@libero.it}

\begin{abstract}
The astronomer E.V. Pitjeva, by analyzing with the EPM2008 ephemerides a large number of planetary observations  including also two years (2004-2006) of normal points from the Cassini spacecraft, phenomenologically estimated a statistically significant non-zero correction to the usual  Newtonian/Einsteinian secular  precession of the longitude of the perihelion of Saturn, i.e. $\Delta\dot\varpi_{\rm Sat} = -0.006\pm 0.002$ \asec; the formal, statistical error is 0.0007 \asec. It can be explained neither by any of the standard classical and general relativistic dynamical effects mismodelled/unmodelled in the force models of the EPM2008 ephemerides
nor by several exotic modifications of gravity recently put forth to accommodate certain cosmological/astrophysical observations without resorting to dark energy/dark matter. Both independent analyses by other teams of astronomers and further processing of larger data sets from Cassini will be helpful in clarifying the nature and the true existence of the anomalous precession of the perihelion of Saturn.
\end{abstract}



\section{Introduction}

At present, the best theory of the gravitational interaction available to us is the Einsteinian General Theory of Relativity (GTR) which has passed so far many observational tests concerning orbital motions and propagation of electromagnetic waves in the (inner) Solar System with excellent results \citep{Wei,Will06,Tur08}. Deviations from the expected behavior have been detected  in the hyperbolic motions of the Pioneer 10/11 spacecrafts after they passed the threshold of approximately 20 Astronomical Units (AU) \citep{And98}, but it is unlikely that the Pioneer anomaly may be ascribed to long-range modifications of the known laws of gravitation \citep{FoP}. The so-called flyby anomaly \citep{And08} consists of a small but unexplained  increase in the velocity of several interplanetary probes (Galileo, NEAR, Rosetta) moving along their hyperbolic orbits at their closest approaches to the Earth; at present, no conventional explanations in terms of known physics have been found. Another anomalous effect which has recently attracted attention is the secular increase of the Astronomical Unit \citep{Kra,Sta04}.
For an overview of such topics see \citep{Lam08}.

In this paper I will focus on a recently detected non-standard feature of the motion of Saturn which, if confirmed as a genuine dynamical effect by further, independent analyses, may be added to the list of the Solar System anomalies not explained by known mundane causes.

 The astronomer E.V. Pitjeva has recently processed a huge data set (1913-2007) of planetary observations of various kinds including also three-dimensional normal point observations of the Cassini spacecraft (2004-2006) with the refined dynamical models of the EPM2008 ephemerides \citep{Pit08a}. They encompass also the action of Eris, the other 20 largest trans-Neptunian objects (TNOs) and a TNOs' massive ring in addition to the usual Newtonian (solar quadrupole mass moment $J_2$, N-body interactions with the major planets, 301 biggest asteroids, massive ring of the small asteroids) and the general relativistic Schwarzschild-like forces of order $\mathcal{O}(c^{-2})$. As a result, she
produced a global solution in which she phenomenologically estimated, among many other parameters, a correction $\Delta\dot\varpi_{\rm Sat}$ to the standard Newtonian/Einsteinian secular, i.e. averaged over one orbital revolution, precession of the longitude of the perihelion $\varpi$ of Saturn whose orbital parameters are in Table \ref{tavola}. It is \citep{Pit08b}
\eqi\Delta\dot\varpi_{\rm Sat} = -0.006\pm 0.002\ {\rm arcseconds\ century}^{-1}\ (''\ {\rm cy}^{-1});\lb{satu}\eqf
it is not compatible with zero at $3-\sigma$ level. Concerning the quoted uncertainty of $0.002$ \asec, it is important to note that it is not the formal  error which is, instead, three times smaller and amounts to $0.0007$ \asec\ \citep{Pit08b}.
There should be considered also the possibility that the realistic uncertainty may be up to 10 times the formal one \citep{Pit08b}, but I believe that until no other independent determinations to be compared with the one of \rfr{satu} will be available, this cannot be decided. Thus, throughout the paper I will rest upon the result of the fit of \rfr{satu}. However, in Section \ref{expla} I will present some considerations, based on the action of the trans-Neptunian objects, disfavoring the possibility that the real uncertainty can be as large as 0.007 \asec.
 $\Delta\dot\varpi_{\rm Sat}$ takes into account, by construction,  any unmodelled/mismodelled dynamical effects affecting the orbit of Saturn.  Previous estimates based on the EPM2006 ephemerides, which did not include  the Cassini data, yielded
\citep{PitPraga}
\eqi \Delta\dot\varpi_{\rm Sat} = -0.92\pm 0.29\ ''\ {\rm cy}^{-1}\ ({\rm formal}\ {\rm error}).\eqf

\begin{table}[!h]
\caption{Orbital parameters of Saturn at the epoch J2000 with respect to the mean ecliptic and equinox of J2000. $a$ is the semimajor axis in AU, $e$ is the eccentricity, $i$ is the inclination of its orbit with respect to the Sun's equator in deg, $P_{\rm b}$ is its orbital period in yr. }\label{tavola}
\centering
\bigskip
\begin{tabular}{cccc}
\hline\noalign{\smallskip}
$a$ (AU) & $e$ & $i$ (deg) & $P_{\rm b}$ (yr) \\
\noalign{\smallskip}\hline\noalign{\smallskip}
$9.53707032$ & $0.05415060$ & $4.6655$ & $29.453$\\
\noalign{\smallskip}\hline
\end{tabular}
\end{table}

Waiting for independent confirmations of \rfr{satu} by other teams of astronomers and further data analysis including, hopefully, more Cassini normal points, in this paper I will address  the following questions. $a)$ May some known standard physical effects, not properly modelled, or unmodelled at all, be the cause of the estimated anomalous retrograde precession of Saturn? $b)$ Could some of the recently proposed modified models of gravity, not modeled in the EPM2008 ephemerides, account for $\Delta\dot\varpi_{\rm Sat}$?

\section{Possible explanations of the anomalous perihelion precession of Saturn}\lb{expla}
In Table \ref{tavola1} I quote the  analytical  expressions and the nominal values of the secular perihelion precessions of Saturn due to the known dynamical effects of classical and relativistic origin, along with some exotic forces recently proposed to explain, e.g., the Pioneer anomaly \citep{And98}, the cosmological expansion without resorting to dark energy \citep{DGP} and the flat rotation curves of galaxies \citep{MOND} without invoking dark matter; also the actions of a spherically symmetric distribution of dark matter \citep{Kri} and  the cosmological constant \citep{Ker} in the Solar System  are considered. It turns out that the majority of the considered effects, modelled or not in the EPM2008 routines, cannot explain both the sign and the magnitude of $\Delta\dot\varpi_{\rm Sat}$ because they induce prograde perihelion precessions. In particular, since the modelling of the trans-Neptunian objects (TNOs) is certainly not yet complete, it may be argued that the uncertainty in their total mass can induce a mismodelled perihelion precession of Saturn large enough to explain \rfr{satu}. Although it is reasonable to assume the modeling of the action of the TNOs  as still preliminary, it seems difficult to attribute the determined anomalous apsidal precession of Saturn to them because the perihelion precession due to them is prograde \citep{KBO}.
Note that if one assumes for the uncertainty in $\Delta\dot\varpi_{\rm Sat}$ the value 0.007 \asec, i.e. 10 times the formal, statistical error, a prograde extra-precession of $+0.001$ \asec\ would be allowed. If attributed to a still imperfectly modeled action of the TNOs, such a positive apsidal precession  would imply $\delta m_{\rm TNOs} = 0.01M_{\oplus}$; it is just the nominal value of the Classical Kuiper Belt Objects (CKBOs) which constitutes about $70\%$ of the entire population of  the TNOs obtained by \citet{Ber04} with non-dynamical techniques\footnote{They used the ACS camera of the Hubble Space Telescope.}. This would be in contrast to the fact that Pitjeva did actually model  the action of the 20 largest TNOs and of a massive ring, i.e. it is difficult to believe that a mismodelling as large as $100\%$ may have occurred given the exquisite level of modeling of the EPM ephemerides.
 The general relativistic gravitoelectric Scwharzschild-like force of order $\mathcal{O}(c^{-2})$ has been modelled in EPM2008. Since the resulting perihelion precession can be written in terms of the standard Eddington-Robertson-Schiff PPN parameters $\beta$ and $\gamma$ \citep{ppn} as
\eqi\dot\varpi_{\rm GE}=\nu\rp{nGM}{c^2 a(1-e^2)}, \eqf with
\eqi\nu = \rp{2+2\gamma -\beta}{3},\eqf
it may be argued that, in principle, $\Delta\dot\varpi_{\rm Sat}$ may be explained in terms of  a deviation $\Delta\nu$ of $\nu$ from its  general relativistic value $\nu_{\rm GTR}=1$.
However, this would imply
\eqi\Delta\nu = -0.428\pm 0.142\eqf which is contradicted by several other independent determinations of $\beta$ and $\gamma$ throughout the Solar System \citep{Wei,Will06,Tur08}.

Concerning the forces able to induce a negative perihelion precession, the Newtonian N-body interactions with the major planets yield the largest retrograde effect; since it is mainly due to the Jupiter, the uncertainty in its mass  might, in principle, induce a mismodelled precession able to accommodate \rfr{satu}. In fact, the answer is negative because the mass of Jupiter is presently known with a relative accuracy of $1\times 10^{-8}$ \citep{Pit08c} which yields a mismodelled precession two orders of magnitude smaller than \rfr{satu}. The retrograde perihelion precession of order $\mathcal{O}(c^{-2})$ due to the general relativistic gravitomagnetic Lense-Thirring force generated by the Sun's rotation \citep{LT}, not modeled in the EPM2008 ephemerides, is smaller than $\Delta\dot\varpi_{\rm Sat}$ by four orders of magnitude.
With regards to the putative exotic forces considered here, the Dvali-Gabadadze-Porrati (DGP) modified gravity \citep{DGP} predicts a negative secular perihelion precession \citep{Lue} of $-0.0005$ \asec; it is too small to explain \rfr{satu} which rules out its existence at $2.75-\sigma$ level. Incidentally, let us note that the positive value of the DGP effect, related to the self-accelerated branch of the cosmic expansion, is ruled out at $3.25-\sigma$ level.
The anomalous acceleration experienced by the Pioneer 10/11 probes at the Saturn orbit \citep{PioSat}, if attributed to a constant and uniform extra-force directed towards the Sun acting also on the planets, induces a retrograde secular precession of the perihelion of Saturn four orders of magnitude larger than \rfr{satu}.
The force quadratic in the radial velocity considered in \citep{Sta08} as a possible explanation of the Pioneer anomaly induces a retrograde precession of the perihelion which, for Saturn, is one order of magnitude larger than \rfr{satu} being incompatible with it at $11-\sigma$ level; in \citep{Ior09} it was still compatible with less recent determinations of $\Delta\dot\varpi_{\rm Sat}$.
Also MOND \citep{MOND} causes  retrograde secular perihelion precessions; however, in the case of Saturn they are either too small or too large  with respect to \rfr{satu} by several orders of magnitude.

It may be argued that some mutual cancelations among different unmodelled/mismodelled effects may have conspired to yield just the estimated value of $\Delta\dot\varpi_{\rm Sat}$, but an inspection of Table \ref{tavola1} shows that this seems to be a very unlikely possibility.

\begin{table}[!h]
\caption{\footnotesize{Nominal values, in \asec, of the secular precessions of $\varpi_{\rm Sat}$ due to known classical and relativistic effects and by some non-standard forces; the effects with $^{\ast}$ have been modelled in EPM2008. The integrated N-body precession can be retrieved at \protect\url{http://ssd.jpl.nasa.gov/txt/p_elem_t1.txt}. I have assumed for the quadrupole mass moment of the Sun $J_2^{\odot} = 2\times 10^{-7}$ \protect\citep{Pir07} and for its  angular momentum $S_{\odot}=190.0\times 10^{39}$ kg m$^2$ s$^{-1}$ \citep{Pij98}.
The precession by the the small asteroid ring has been computed according to \protect\citet{Fie} with $m_{\rm ring} = 0.34\times 10^{-10}\ M_{\odot}$ and $r_{\rm ring}=2.8$ AU.
TNOs are the trans-Neptunian objects modelled as a ring of mass $m_{\rm TNOs}$ and inner and outer radii $R_{\rm min}$ and $R_{\rm max}$, respectively \citep{KBO}. DGP is the \protect\citet{Lue} perihelion precession in the Dvali-Gabadadze-Porrati \citep{DGP} multidimensional braneworld model;
$r_0 \approx 5$Gpc. Pioneer  (Saturn) is the Pioneer anomaly at the Saturn's orbit $A_{\rm Pio}=-(1.8\pm 6.4)\times 10^{-10}$ m s$^{-2}$ \citep{PioSat}. Pioneer $(|v_r|)$  and Pioneer $(v_r^2)$ are the velocity-dependent forces proposed in \protect\citep{Sta08}; for them I used $\mathcal{K}=7.3\times 10^{-14}$ s$^{-1}$ and $\mathcal{H} = 6.07\times 10^{-18}$ m$^{-1}$. The effect of a spherically symmetric distribution of dark matter has been worked out in, e.g., \citep{Kri}, while that of MOND is due to \citep{Ser06} with $r_{\rm MOND}=\sqrt{GM_{\odot}/A_0}$, $A_0=1.2\times 10^{-10}$ m s$^{-2}$. The apsidal precession induced by the cosmological constant $\Lambda\approx 10^{-52}$ m$^{-2}$ is due to \citet{Ker}. The other parameters used are $a,e,i$, semimajor axis, eccentricity and inclination, respectively, of the planetary orbit, $n=\sqrt{GM_{\odot}/a^3}$ Keplerian mean motion,  $M_{\odot},R_{\odot}$ mass and equatorial radius, respectively, of the Sun, $G$ Newtonian gravitational constant, $c$ speed of light in vacuum.} }\label{tavola1}
\centering
\bigskip
\footnotesize{\begin{tabular}{ccc}
\hline\noalign{\smallskip}
Dynamical effect & Analytical expression   & $\dot\varpi$ (\asec) \\
\noalign{\smallskip}\hline\noalign{\smallskip}
N-body$^{\ast}$ & numerical integration  & $-1508.313$ \\
Solar quadrupole$^{\ast}$ $J_2^{\odot}$ & $\rp{3}{2}\rp{n J_2}{(1-e^2)^2}\left(\rp{R_{\odot}}{a}\right)^2\left(1-\rp{3}{2}\sin^2 i\right)$ & $3\times 10^{-7}$\\
 Small asteorid ring$^{\ast}$  & $\rp{3}{4}\sqrt{\rp{G}{M_{\odot} a^7} }\rp{m_{\rm ring}r^2_{\rm ring}}{1-e^2}$ & $1\times 10^{-5}$ \\
TNOs$^{\ast}$ & $\rp{3}{4}\sqrt{\rp{Ga^3(1-e^2)}{M_{\odot}}}\rp{m_{\rm TNOs}}{(R_{\rm max}+R_{\rm min})R_{\rm max}R_{\rm min}}$ & $>0$\\
Schwarzschild$^{\ast}$ & $\rp{3nGM_{\odot}}{c^2 a (1-e^2)}$ & 0.014\\
Lense-Thirring & $-\rp{4GS_{\odot}}{c^2 a^3 (1-e^2)^{3/2}}$ & $-1\times 10^{-7}$\\
DGP & $\mp\rp{3c}{8r_0}$ & $\mp 0.0005$\\
%
%
Pioneer (Saturn) & $A_{\rm Pio}\sqrt{\rp{a(1-e^2)}{GM_{\odot}}}$& $-12.130\pm 43.130$\\
Pioneer $(|v_r|)$ & $-\rp{\mathcal{K}\sqrt{1-e^2}}{\pi}\left[\rp{2e-(1-e^2)\ln\left(\rp{1+e}{1-e}\right)}{e^2}\right]$ & 1.091\\
Pioneer $(v_r^2)$ & $\rp{\mathcal{H}na\sqrt{1-e^2}}{e^2}\left(-2 + e^2 + 2\sqrt{1-e^2}\right)$ & $-0.028$\\
Dark matter & $\rp{4\pi G\rho_{\rm dm}\sqrt{1-e^2}}{n}$ & $>0$\\
MOND ($k_0=1/2$, $m=2$) & $-k_0 n\left(\rp{a}{r_{\rm MOND}}\right)^{2m}m$ & $-1\times 10^{-5}$\\
MOND ($k_0=1$, $m=1$) & $-k_0 n\left(\rp{a}{r_{\rm MOND}}\right)^{2m}m$ & $-8.098$\\
Cosmological constant & $\rp{1}{2}\left(\rp{\Lambda c^2}{n}\right)\sqrt{1-e^2}$ & $4\times 10^{-13}$\\

\noalign{\smallskip}\hline
\end{tabular}
}
\end{table}

The detected anomalous perihelion precession of Saturn may be used to phenomenologically constrain  the existence of an
unknown constant and uniform acceleration directed towards the Sun continuously existing in the spatial regions swept by the Saturn's orbital motion during the time interval spanned by the data set used (1913-2007) covering about four orbital revolutions of Saturn.
It is
\eqi A_{\rm Sat} = -(9.14 \pm 3.04)\times 10^{-14}\ {\rm m}\ {\rm s}^{-2}.\lb{accel}\eqf
However, its existence in the inner regions of the Solar System is ruled out by the estimated corrections to the Newtonian-Einsteinian perihelion precessions of Venus, Earth and Mars, as shown by Table \ref{tavola2}.
\begin{table}[!h]
\caption{First row: estimated corrections to the standard perihelion precessions of Mercury  \protect\citep{Pit05}, Venus \protect\citep{Pit08d},  Earth  \protect\citep{Pit05}  and Mars  \protect\citep{Pit05}, in \asec. The quoted errors are realistic.  Second row:
anomalous perihelion precessions, in \asec, of Mercury, Venus,  Earth and Mars induced by \rfr{accel}. }\label{tavola2}
\centering
\bigskip
\begin{tabular}{cccc}
\hline\noalign{\smallskip}
Mercury & Venus & Earth & Mars \\
\noalign{\smallskip}\hline\noalign{\smallskip}
$-0.0036 \pm 0.0050$ & $-0.0004\pm 0.0005$  & $-0.0002\pm 0.0004$   & $0.0001\pm 0.0005$ \\
$-0.0011$ & $-0.0017$ & $-0.0019$ & $-0.0023$\\
\noalign{\smallskip}\hline
\end{tabular}
\end{table}
Concerning \rfr{accel}, it must be noted that it could not be reproduced by a Yukawa-like term
\eqi A_{\rm Yuk} = -\rp{GM\alpha}{r^2}\left(1+\rp{r}{\lambda}\right)\exp\left(-\rp{r}{\lambda}\right)\lb{yuki}\eqf evaluated at the Saturn's orbit.
Indeed, in \citep{YUK} it has been shown that the estimated corrections to  the standard precessions of the perihelia of the inner planets
constrain $\alpha$ and $\lambda$ to $\alpha\leq 4\times 10^{-11}$, $\lambda\leq 0.18$ AU;  such values in \rfr{yuki} yield for Saturn
\eqi A_{\rm Yuk} = -2\times 10^{-15}\ {\rm m}\ {\rm s}^{-2}.\eqf
On the other hand, typical values for $\alpha$ and $\lambda$ able to fit astrophysical observations of distant galaxies, i.e. $\alpha = -3\times 10^{-8}$, $\lambda = 33,000$ AU  \citep{Mof},
would yield for Saturn
\eqi A_{\rm Yuk} = 1.98\times 10^{-12}\ {\rm m}\ {\rm s}^{-2}.\eqf

\section{Discussion and conclusions}
Based on the analysis presented, summarized by Table \ref{tavola1},  I conclude that the recently estimated anomalous retrograde apsidal precession of Saturn  $\Delta\dot\varpi_{\rm Sat} = -0.006\pm 0.002$ \asec\ cannot be explained by any of those standard Newtonian and Einsteinian dynamical effects which have  been mismodeled  (or unmodeled at all) in the force models of the EPM2008 ephemerides. The same holds also for many exotic modifications of gravity proposed in the recent past to explain various kinds of cosmological/astrophysical observations. In particular, the DGP braneworld model is ruled out at about $3-\sigma$ level, while the existence of the force quadratic in the radial velocity proposed to explain the Pioneer anomaly must be excluded, at least in the spatial regions swept by the Saturn's orbit, at $11-\sigma$ level. Table \ref{tavola1} shows also that a finely-tuned cancelation of several unmodelled/mismodelled effects yielding just the estimated $\Delta\dot\varpi_{\rm Sat}$ is unlikely.
Both independent analyses by other teams of astronomers will be important in order to clarify the nature and the genuine existence of the anomalous behavior of the Saturnian perihelion. Moreover, it will be important for further studies to include  the largest number of normal points as possible from spacecraft-based missions-in particular Cassini-covering the largest portion as possible of a full orbital revolution of Saturn.

\section*{Acknowledgments}
I gratefully thank E.V. Pitjeva (Institute of Applied Astronomy, Russian Academy of Sciences, St. Petersburg) for her latest determinations of the orbital motion of Saturn.
I am the sole responsible for the interpretation of the data presented here and it does not reflect the official position of any other person or institution.


\end{document}